# A HIGH ENERGY LHC MACHINE. EXPERIMENTS 'FIRST' IMPRESSIONS

M. Nessi, CERN, Geneva, Switzerland

*Abstract*

These days, while the landscape of discoveries at LHC has yet to be unveiled, planning for upgrades twenty years or more in advance towards a possible experimental scenario, might sound very imaginative and ambitious. Nevertheless, as plans are being worked out for the High Luminosity LHC upgrade, it is possible to plan keeping the ATLAS and CMS detectors operational for the following High Energy phase. The natural and radiation-induced aging of some components, calorimeters especially, needs to be carefully addressed. Even planning for a very new detector might not be unreasonable.

## INTRODUCTION

Trying to extrapolate a possible experimental scenario twenty years or more in advance, might sound very ambitious and imaginative, in particular today before knowing the discovery landscape of the present LHC.

At some point, while scanning through the possible rare physics signals, luminosity at LHC will not buy more statistics. Cross-sections will become simply too small and will drop by many orders of magnitude, in particular as a function of mass. Energy will buy much more, because rare physics cross-sections, and in particular if large mass objects are involved, will be boosted by the larger amount of energy available to create heavy objects. We assume that this possible changeover of strategy between high luminosity and high energy will become interesting around 2000-3000 $fb^{-1}$ of collected integrated luminosity. At that time, probably around 2030-2032, both multi-purpose detectors, ATLAS [1] and CMS [2], will be still operational.

## POSSIBLE DETECTOR REQUIREMENTS

In today's scenario, while no discoveries have been announced yet, the kind of physics we will be exploring at high energy will basically be the same we are investigating now (see Fig 1.), but with some "nuances".

- Discovery of high mass new particles (beyond what will be explored at the HL-LHC, m ~ 2.5 TeV).
- Precision measurements of known Standard Model physics (heavy flavors precision measurements and rare decays).
- Measuring in detail properties of newly discovered phenomena (masses and couplings of sparticles as an example).
- Precision measurements of LHC discoveries (Higgs spin, self- couplings, rare decays, ...).
- Searches for new phenomena, not anticipated by theory.

It is therefore hard to guess which parameters in today's detector properties might be relaxed. Today probably none. If we will still be looking for SUSY-type phenomena, with large multiplicities of leptons, jets and heavy flavor decays and missing transverse energy, then the detectors will have to count on:

- Lepton identification (in particular electrons versus jets), photon and muon identification.
- b and c quark decay tagging, via secondary vertex tagging.
- Excellent missing energy resolution, which implies detector coverage down to large pseudo-rapidity values.
- Excellent calorimeters performance in terms of resolution and energy scale.
- Excellent tracking efficiency (>98%).

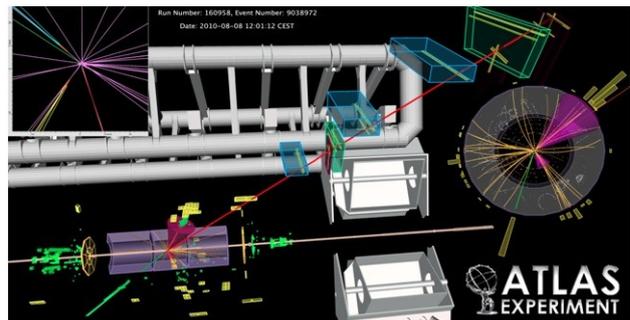

Figure 1: Example of a recently recorded top event in ATLAS, showing the various detectors components involved.

On top of this, and after a few years of enthusiasm for the high-energy regime, the community will certainly ask (because of the positive experience at HL-LHC) to run with high luminosity too. This will reopen the issue and stress even further the detector requirements.

- A high number of pile-up events with many tracks and a large risk of fake hits/tracks association (see Fig. 2).
- An important cavern background, in particular from slow neutrons captured in the detector materials.
- Unprecedented levels of radiation and track densities, in particular in the forward detectors, that will limit their effectiveness.

It is therefore impossible today to assume that some of the present detector properties or requirements will be relaxed. This means that presently we have to assume that in 20 years from now we will be able to operate and maintain the existing detectors as we do today, after an important upgrade of the innermost components which we are planning for the HL-LHC.

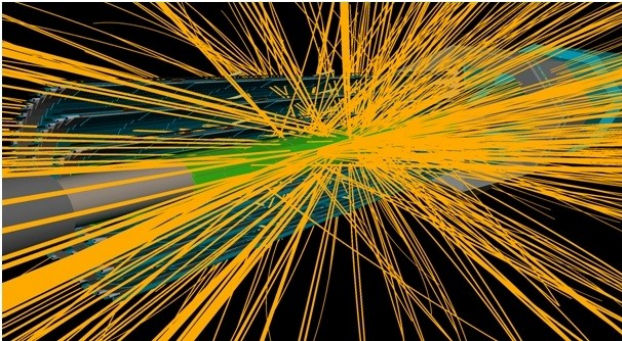

Figure 2: High tracks density in the vertex detector with just 50 pile-up events associated.

## DETECTORS CONCERNS

After the HL-LHC experience, the detectors will be old in their structure and constituent materials. From the time of construction (1996-2008) about 30 years will have elapsed. Some more critical parts like rubber components, O-rings, PCBs, cables and connectors, optical fibers, cryo- and vacuum infrastructure will need a careful analysis and probably will need to be replaced.

A large part of the electronics (front-end and back-end) will be obsolete and no longer possible to keep operational. The procurement of electronics spare components will be an issue.

Some components will have been heavily irradiated. The innermost parts will be already classified as potential nuclear waste. Access will be very limited in the regions around the beam pipes ($\sim$ 2 m radius) and near to the TAS. The main issue will be the irradiation of services and electronics. In the region around the beam pipe we will probably be at the level of a few mSv/h. Today, running at a peak luminosity of 3 to $5 \times 10^{31}$ cm$^{-2}$ s$^{-1}$, we observe online an activation level around the ATLAS beam pipe well in line with the calculations obtained by simulation.

Activation and radioactive contamination, and Radio Protection (RP) issues in general will become fundamental from 2016 on, and on the very long term they will represent a real problem. We have in any case to change our culture and be more proactive in this domain.

### Inner Detectors

For the HL-LHC both collaborations will have constructed a new inner detector with very high granularity and with radiation-harder sensors and front-end electronics (~2020). R&D on a new generation of Silicon- or Diamond-based sensors has already started. For example, 3D Silicon strip detectors represent a very promising technology if the industrialization process will be effective (see Fig. 3).

Having in mind to use the same layout for the HE phase, one has now to introduce in today's upgrade requirements the possibility to upgrade and exchange inner detector (ID) components continuously as a function of time. This is particularly true and valid for the innermost layers (b-layer and pixel detector in general). We have also to add to our 2020 specifications a radiation resistance up to ~6000 fb$^{-1}$. The alternative is to assume a new upgrade of the entire inner detectors in the early thirties, as we will have done for the HL-LHC.

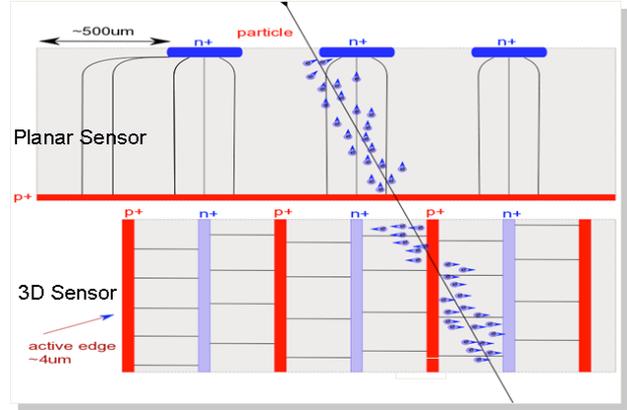

Figure 3: principle of function of planar versus 3D Silicon strip detectors. In the 3D case, the drift time, as well as the active edge dimension, are reduced considerably.

### Calorimeters

Calorimeters using scintillating dopants risk being completely irradiated and therefore will have changed their transparency property regarding optimum light collection. No idea about what to do in such a case. Especially in the ATLAS case, it is practically impossible to extract and replace the calorimeters without dismounting most of the detector. This might be the critical problem we will have to face. Either one accepts a reduced light collection and a bigger constant term in the energy resolution, or one needs to start replacing components. All present light detectors (photomultipliers or diodes) will have reached the end of their life cycle and will need to be replaced.

- For example the ATLAS Tile calorimeter performance will need to be evaluated as a function of radiation. Its injection-molded 40 tons of scintillator might be fully aged in its properties and compromised by radiation. No way to dismount it without dismounting a major part of ATLAS. Maybe something can be done in the end-caps. If not, one has just to accept a reduced performance.
- Similar reasoning for the CMS calorimeters (crystals + hadronic scintillators). In particular in the end-cap regions, radiation will compromise the crystals light transmission, probably to a point where crystals need to be replaced. Differently from ATLAS, here the access is simpler and it might be easier to replace end-caps components (i.e. crystals)
- The ATLAS LAr calorimeter will be very radioactive and will be polluted with material, which one can consider as dust and might be a source of electrical shorts in the electrodes, producing HV breakdowns. Here solutions have envisaged on how to solve the problem if one comes to a showstopper. Such an intervention will require opening the cryostats underground to gain access to the active components (probably just in the end-caps). The intervention will be very difficult, because of the radiation levels and will require at least 3 years of downtime.

*Muon Spectrometers and Magnets*

All experimental magnets, should still be fully operational. Over the years the operating fields may have been increased by 10-15%, increasing the resolution capability to the trackers. The controls and all peripheral services will be obsolete in their technology. An effective upgrade will be easy. Probably it will happen already in the mid twenties.

For the muon spectrometer (trigger and precision chambers) the problem lies in the natural aging of the critical components and of the base materials in general. Most of the active components have been designed for a lifetime of 15-20 years. These are gaseous detectors, therefore less robust and more subject to stresses in terms of mechanics and services (gas leaks, gas distribution infrastructure, connectors, resistive materials,…).

Already for the High Luminosity upgrades we foresee to replace the end-cap chambers in the high rapidity regions with more granular and trigger-effective components. New technologies will be adopted. In the same spirit, it is likely possible to start replacing around 2030 most or all of the muon stations during regular shutdowns. In the case of ATLAS, for some chambers a direct replacement will not be possible, access being the problem. An unconventional approach will be needed.

As for the ID, the muon spectrometer strategy should be a continuous upgrade over time, profiting from all shutdowns of the LHC machine, while keeping the technology up to date.

## A NEW DETECTOR

Why not to think and plan for a very new detector in general, in parallel to ATLAS and CMS?

If we go the HE-LHC way, probably it means no Linear Collider for a while! A large detector community preparing today already for the Linear Collider is in standby, with plenty of new ideas and several new technologies to be deployed.

A new detector could be tuned from the beginning to the type of new discoveries the LHC will make and go beyond in a more effective way. It will take 16-18 years to achieve a fully functional new detector, and this means that a green light to move in this direction should be given around 2015. A new detector might imply new civil engineering work to prepare a new experimental cavern in today's LHCb or ALICE location.

## CONCLUSIONS

Thinking about the ATLAS and CMS evolution in the HE-LHC scenario, the following arguments might apply:

- Most of the electronics will need to be rebuilt and upgraded. This will partially happen already for the HL-LHC, leaving therefore no reason not to do it later as well. This would solve the problem of obsolete technologies.
- Inner detectors will be upgraded after 2020, and there is no reason not to continue doing it further, maybe just in a modular way. The story is similar for the muon spectrometer. Consolidation/upgrade can be continuous.
- The calorimeters are the more critical items, needing a particular evaluation, possibly representing a serious showstopper.
- Over time, the trigger hardware and strategies will be revisited. Doing this already for the HL-LHC. Physics will guide us!

An experimental program based only on the existing detectors might be risky, also giving the fact that the investments needed for a new LHC machine with more energy will be substantial. Planning for a fully new detector might be a more rational approach. It might take more time to conceive a new detector than to upgrade the accelerator. Thus, a strategical decision might due in just a few years from now.